\documentclass[%
 preprint,
superscriptaddress,
 amsmath,amssymb,
 aps,
prb,
]{revtex4-1}

\usepackage{graphicx}
\usepackage{dcolumn}
\usepackage{multirow}
\usepackage{bm}

\begin{document}

\title{Spin-polarized ballistic conduction through correlated \\ Au--NiMnSb--Au heterostructures}

\author{C. Morari}
\affiliation{National Institute for Research and Development of Isotopic and
Molecular Technologies, 67-103 Donat,  400293 Cluj Napoca, Romania}
\author{W. H. Appelt}
\affiliation{Theoretical Physics II,
Institute of Physics, University of Augsburg, 86135 Augsburg, Germany}
\address{Augsburg Center for Innovative Technologies, University of Augsburg,
86135 Augsburg, Germany}
\author{A. Prinz-Zwick}
\author{U. Eckern}
\affiliation{Theoretical Physics II,
Institute of Physics, University of Augsburg, 86135 Augsburg, Germany}
\author{U. Schwingenschl\"ogl}
\address{PSE Division, KAUST, Thuwal 23955-6900, Kingdom of Saudi Arabia}
\author{A. \"Ostlin}
\affiliation{Theoretical Physics III, Center for Electronic
Correlations and Magnetism, Institute of Physics, University of
Augsburg, 86135 Augsburg, Germany}
\author{L. Chioncel}
\affiliation{Augsburg Center for Innovative Technologies, University of Augsburg,
86135 Augsburg, Germany}
\affiliation{Theoretical Physics III, Center for Electronic Correlations and
Magnetism, Institute of Physics, University of Augsburg, 86135 Augsburg,
Germany}

\begin{abstract}
We examine the ballistic conduction through Au--NiMnSb--Au heterostructures consisting of up to four units of NiMnSb in the scattering region.
We investigate the dependence of the transmission function computed within the local spin density approximation (LSDA) of the density functional theory (DFT) on the number of half-metallic units in the scattering region.
For a single NiMnSb unit the transmission function displays a spin polarization of around 50\% in a window of $1\,$eV centered around the Fermi level. 
By increasing the number of layers an almost complete spin polarization of the transmission is obtained in the same energy window.
Supplementing the DFT-LSDA calculations with local electronic interactions, of Hubbard-type on the Mn sites, leads to a hybridization between the interface and many-body states. 
The significant reduction of the spin polarization seen in the density of states is not apparent in the spin-polarization of the conduction electron transmission, which suggests the localized nature of the hybridized interface and many-body induced states.
\end{abstract}

\pacs{Valid PACS appear here}
\maketitle

\section{Introduction}
Multi-layered heterostructures composed of alternating magnetic and non-magnetic metals offer large flexibility in tailoring spin-sensitive or spin-contrasted electron transport properties of spintronic devices. 
Highly spin-polarized materials such as half-metallic ferromagnets (HMF) are expected to play a crucial role~\cite{gr.mu.83,ka.ir.08}. The extreme spin-polarization of half-metals (i.e., $100\%$) is a consequence of  their band-structure: these materials are metallic for one spin channel, and insulating or semiconducting for the other one. The prototype half-metallic material is the semi-Heusler compound NiMnSb~\cite{gr.mu.83}.

High quality films of NiMnSb alloys have  been grown by molecular epitaxy~\cite{ro.bo.00}, or magnetron sputtering~\cite{kw.sa.16}, the measured conduction electron spin polarization was found to be smaller than about 58$\%$~\cite{so.by.98}. 
This polarization value is consistent with the small perpendicular magnetoresistance measured for NiMnSb in a spin-valve structure, and a superconducting and a magnetoresistive tunnel junction~\cite{ka.te.90}.
It was shown that during the growth of the NiMnSb thin films, first Sb and then Mn atoms segregate to the surface, decreasing the spin polarization~\cite{ri.no.00}.
By removing the excess Sb a nearly stoichiometric ordered alloy surface terminated by a MnSb layer is obtained with a spin polarization of about $67 \pm 9 \%$ at room temperature~\cite{ri.no.00,bo.ko.00}. 
Magnetic circular dichroism measurements show a reduction of both the manganese and nickel moments around $80\,$K. 
Borca et al.~\cite{bo.ko.01} concluded that at this specific temperature, a transition from a half-metallic into a normal ferromagnetic state takes place and the magnetic coupling of the manganese and nickel moments is lost.
The computational study of Le\v{z}ai\'c et al.,~\cite{le.ma.06} suggests a vanishing of the Ni moment at around $80\,$K with a simultaneous loss of polarization. 
A similar study has been performed including interfaces~\cite{bo.ko.00,bo.ko.01}. 
The $80\,$K anomaly, however, is not reflected in the spontaneous magnetization of bulk NiMnSb~\cite{ot.fe.87}; neither are experimental results known for the $80\,$K anomaly at HMF interfaces. 
An alternative scenario which addresses the contradiction between the theoretical predictions and the experimental results concerning the spin polarization in half-metals  is provided by finite-temperature effects~\cite{skom.07} and non-quasiparticle states~\cite{ch.ka.03,ch.ar.06,ch.ar.09,ka.ir.08}.

Advanced materials with high-performance half-metallicity are desired for further improvement of current-perpendicular-to-plane giant magnetoresistance (CPP-GMR) devices. 
Recent experiments~\cite{we.ku.15} have been conducted on (001)-oriented fully epitaxial NiMnSb heterostructures with Ag spacer layers. 
Negative anisotropic magnetoresistance (AMR) ratio and  small discrepancies of the AMR amplitudes between room temperature and $10\,$K were observed in a single epitaxial NiMnSb film, indicating robust bulk half-metallicity against thermal fluctuations in the half-Heusler compound. The modest CPP-GMR ratios  were attributed to interface effects between NiMnSb and Ag.
Gold is a frequently used material for the lead setup in  transport computations, and it has a similar lattice constant ($\approx 4.08\,$\AA) as silver ($\approx 4.09\,$\AA). 
Therefore we consider in our ballistic transport setup a NiMnSb(001) to Au(001) interface which shows a lattice mismatch of $\approx 2\,\%$ so that fully epitaxially grown heterostructures are likely to show little stress at the interface. 

In this paper we address the question of ballistic conduction by examining the density of states and transmission in the vicinity of the Fermi energy ($E_F$).
It is of interest to investigate under which conditions
the scattering region involving NiMnSb units can exhibit half-metallic properties and ballistic transport in the direction of growth.
For a considerable number of bulk half-metallic materials~\cite{ch.ka.03,ka.ir.08}, the interaction induced non-quasiparticle states located within the half-metallic gap were proven to significantly reduce the conduction electron spin-polarization, while no significant change in the magnetic moment takes place.
We extend the study on the existence  of non-quasiparticle states at interfaces and investigate their possible impact on steady-state transport.

The article is organized as follows: Sec.~\ref{sec:method} gives a brief review of the methodology. 
The standard equations to compute transport properties are presented in   Sec.~\ref{sec:ball_lda}. 
The extension including correlations, in the spirit of dynamical mean field theory (DMFT), is presented in Sec.~\ref{sec:ball_dmft}.
We review also the main ideas of the perturbative SPTFLEX-solver~\cite{ka.li.02,li.ka.97,ch.vi.03} of DMFT. 
In Sec.~\ref{sec:results}, we discuss the results for the density of states (DOS) and the transmission in the presence of electronic interactions. Sec.~\ref{sec:conc} provides the conclusion.

\section{Methods}
\label{sec:method}
We use the ``two-step'' approach presented in our previous paper~\cite{ch.mo.15}, in which the Landauer transmission probability is calculated within the {\sc smeagol} non-equilibrium Green's function (NEGF) based electron transport code~\cite{ro.su.06,ro.su.05,ru.sa.08}. 
The {\sc smeagol} imports the DFT Hamiltonian from the {\sc siesta} code~\cite{so.ar.02}, which uses pseudopotentials and expands the wave functions of valence electrons over the basis of numerical atomic orbitals (NAOs).
In the original paper~\cite{ch.mo.15}, the many-body corrections to the Green's function were evaluated using  DMFT~\cite{me.vo.89,ko.vo.04,held.07} in an exact muffin-tin orbitals (EMTO)-based package~\cite{an.sa.00,vi.sk.00,vito.01}, which uses a screened KKR approach~\cite{wein.90}. 
These corrections were then passed to {\sc smeagol} for the calculation of the transmission matrix for ballistic transport throughout the heterostructure.
In the present paper, we instead of the EMTO method use the full-potential linearized muffin-tin orbitals (FPLMTO) method, as implemented in the {\sc RSPt} code~\cite{rsptbook,gr.ma.12}.
The FPLMTO method makes it possible to go beyond shape approximations and treat the full potential, while still keeping a minimal physical basis set.
Self-consistent DFT calculations are performed separately in {\sc smeagol} and in the {\sc RSPt} code. 
The many-body self-energy is then evaluated after self-consistency in the {\sc RSPt} code, and passed to the {\sc smeagol} Green's function to compute the transmission according to the Landauer-B\"uttiker formalism.~\cite{land.57b,land.88,butt.86,butt.88}

\subsection{Ballistic transport for electronic systems}
\label{sec:ball_lda}
The electronic transport through a device can be addressed  in the Landauer-B\"uttiker formulation~\cite{land.57b,land.88,butt.86,butt.88}. According to this model  the current flow through a device is considered as a transmission process across a finite-size scattering region placed between two semi-infinite leads, connected, in their turn (at infinity), to charge reservoirs.
The  quantity of interest is the conductance, which, within linear response, is
given by:
\begin{eqnarray}
\mathcal{G} &=& \frac {e^2}{h}\frac{1}{\Omega_\mathrm{BZ}}\sum_{\sigma=\uparrow,\downarrow}\int_\mathrm{BZ}\!d\mathbf{k}_\parallel T_{\sigma}(\mathbf{k}_\parallel,E_F), \label{conductance} \\
T_{\sigma}(\mathbf{k}_\parallel,E) &=& {\rm Tr} \left[
\mathbf{\Gamma}_{L}^\sigma (\mathbf{k}_\parallel,E) \mathbf{G}^{\sigma\dagger} (\mathbf{k}_\parallel,E)
\mathbf{\Gamma}_{R}^\sigma (\mathbf{k}_\parallel,E) \mathbf{G}^{\sigma}(\mathbf{k}_\parallel,E)\right], \label{landauer} 
\end{eqnarray}
where $e$ is the electron charge, $h$ is the Planck constant, $e^2/h$
is half the quantum of conductance, and $T_\sigma(\mathbf{k}_\parallel,E_F)$
is the spin-dependent transmission probability from one lead to the other
for electrons at the Fermi energy with the transverse wave-vector
$\mathbf{k}_\parallel$ perpendicular to the current flow.
The integral over $\mathbf{k}_\parallel$ goes over the Brillouin zone (BZ) perpendicular to the transport direction, and $\Omega_\mathrm{BZ}$ is the area of the BZ. 
The retarded  Green's function $\mathbf{G}^{\sigma}(\mathbf{k}_\parallel,E)$ has the following form:
\begin{equation}
\mathbf{G}^\sigma(\mathbf{k}_\parallel,E) = \left[
\epsilon^+\mathbf{S}(\mathbf{k}_\parallel)
-\mathbf{H}^\sigma(\mathbf{k}_\parallel)
-\mathbf{\Sigma}_{L}^\sigma(\mathbf{k}_\parallel,E)
-\mathbf{\Sigma}_{R}^\sigma(\mathbf{k}_\parallel,E)
\right]^{-1}.
\label{Green_add}
\end{equation}
All terms presented are matrices $[\mathbf{G}^\sigma(\mathbf{k}_\parallel,E)]_{\mu\nu}$, labelled by the global indices  $\mu ,\nu$ which run through the basis functions at all atomic positions in the scattering region. $\mathbf{S}(\mathbf{k}_\parallel)$ represents the orbital overlap matrix, and the energy shift into the complex plane, $\epsilon^+ = \lim_{\delta \to 0^{+}}(E + i \delta)$, has been introduced to respect causality. 
$\mathbf{H}^{\sigma}(\mathbf{k}_\parallel)$ is the Hamiltonian of the scattering region for spin $\sigma$; the right and left self-energies
$\mathbf{\Sigma}_{R}^\sigma(\mathbf{k}_\parallel,E)$ and
$\mathbf{\Sigma}_{L}^\sigma(\mathbf{k}_\parallel,E)$ describe the energy-,
momentum- and spin-dependent hybridization of the scattering region with
the right and left leads, respectively \cite{ru.sa.08}. 
Therefore, $\mathbf{G}^\sigma(\mathbf{k}_\parallel,E)$  is formally the retarded Green's function associated to the effective, non-Hermitian Hamiltonian
$\mathbf{H}^\sigma_{\mathrm{eff}}(\mathbf{k}_\parallel,E)=\mathbf{H}^\sigma(\mathbf{k}_\parallel) -\mathbf{\Sigma}_{L}^\mathbf{\sigma}(\mathbf{k}_\parallel,E) -\mathbf{\Sigma}_{R}^\sigma(\mathbf{k}_\parallel,E)$. In Eq.~(\ref{landauer}),
$\mathbf{\Gamma}_{L(R)}^\mathbf{\sigma}(\mathbf{k}_\parallel,E)=i\big[\mathbf{\Sigma}_{L(R)}^\sigma(\mathbf{k}_\parallel,E)-\mathbf{\Sigma}_{L(R)}^{\sigma\dagger}(\mathbf{k}_\parallel,E)\big]$
is the so-called left (right) broadening matrix that accounts for the hybridization-induced broadening of the single-particle energy levels of the scattering region. 
Importantly, for non-interacting electrons, it has been proven that the Landauer and the Kubo approaches are equivalent~\cite{fi.pa.81}, so that the linear-response transport properties of a system can be computed with either formalism. The Landauer approach has been systematically applied in conjunction with DFT in order to perform calculations of the conductance of different classes of real nano-devices~\cite{ta.gu.01}. In this approach the DFT provides a single-particle theory in which the Kohn-Sham eigenstates are interpreted as single-particle excitations. Although this is only valid approximately, DFT-based transport studies have provided insightful results concerning the role of the band-structure in the electron transport process through layered heterostructures \cite{sc.ke.95,sc.ke.98,bu.zh.01,ru.mr.09,ca.ar.12}.

\subsection{Ballistic transport for correlated electrons}
\label{sec:ball_dmft}

In a multilayer heterostructure the dimensionality of the problem  
requires a layer-resolved DMFT~\cite{me.vo.89,ge.ko.96,ko.vo.04} solution for the correlated problem. Therefore, the setup we consider involves a self-consistent calculation for the heterostructure to include electron-electron interaction beyond the LDA or generalized gradient approximation (GGA) of the DFT explicitly. Accordingly, the retarded Green's function of Eq.~(\ref{Green_add}) has to be modified. 
We include electron-electron interactions in the form of a multi-orbital local Hubbard term 
$\frac{1}{2}\sum_{{i \{m, \sigma \} }} U_{mm'm''m'''} c^{\dag}_{im\sigma}c^{\dag}_{im'\sigma'}c_{im'''\sigma'}c_{im''\sigma}$ within the interacting region. Here, $c_{im\sigma}$($c^\dagger_{im\sigma}$) destroys (creates) an electron with spin $\sigma$ on orbital $m$ at the site $i$. The Coulomb matrix elements $U_{mm'm''m'''}$  are expressed in the standard way~\cite{im.fu.98} in terms of three Kanamori parameters $U$, $U'$ and $J$.  
The interaction is treated in the framework of DMFT~\cite{ko.vo.04,ko.sa.06,held.07}, with a spin-polarized T-matrix Fluctuation Exchange (SPTF) type of impurity solver~\cite{ka.li.02} implemented within the FPLMTO basis set~\cite{rsptbook,gr.ma.12}.
The SPTF approximation is a multiband spin-polarized generalization of the fluctuation exchange approximation (FLEX)~\cite{bi.sc.89,ka.li.99}. 
In the context of lattice models it describes the interaction of quasiparticles with collective modes. In practice, it is a perturbative expansion of the self-energy in powers of $U$, with a resummation of specific classes of diagrams, such as ring diagrams and ladder diagrams. 
The expansion remains reliable when the strength of the interaction $U$ is smaller than the bandwidth of the material. This is a valid approach for NiMnSb as its bandwidth is about $\approx 8$ eV, and the relevant values for the local Coulomb parameter are in the range of $U \approx 2\dots 3$ eV~\cite{ch.ka.03}. 
Justifications, further developments, and details of this scheme can be found in Ref.~\onlinecite{ka.li.02}.

For the case of half-metallic ferromagnets it was demonstrated~\cite{ka.ir.08} by model as well as realistic electronic structure calculations that many-body effects are crucial for half-metals: they produce states with tails that cross the Fermi level so that the gap is closed and half-metallicity is lost~\cite{ch.ka.03,al.ch.10,ch.mo.15,ch.ka.05,ch.ar.06,ch.ar.09}.
The origin of these many-body non-quasiparticle (NQP) states is connected with ``spin-polaron'' processes: the spin-down low-energy electron excitations, which are forbidden for the HMF in the
one-particle picture, turn out to be possible as superpositions of
spin-up electron excitations and virtual magnons~\cite{ed.he.73,ka.ir.08}.
Spin-polaron processes are described within the SPTF approach by the fluctuation potential matrix $W^{\sigma \sigma ^{\prime }}(i\omega_n)$ with $\sigma=\pm$, defined as follows~\cite{ka.li.99}:
\begin{equation}\label{W}
{\hat W}(i\omega )=\left(
\begin{array}{cc}
{W}^{++}(i\omega_n ) & {W}^{+-}(i\omega_n ) \\
{W}^{-+}(i\omega_n ) & {W}^{--}(i\omega_n )
\end{array}
\right).
\end{equation}

The essential feature here is that the potential (\ref{W}) is a complex energy-dependent matrix in spin space with {\it off-diagonal} elements:
\begin{equation}
W^{\sigma, -\sigma}(i\omega_n)=U^m(\chi^{\sigma, -\sigma}(i\omega_n )-
\chi_0^{\sigma, -\sigma}(i\omega_n ))U^m,
\end{equation}
where $U^m$ represents the bare vertex matrix corresponding to the transverse magnetic channel, $\chi^{\sigma, -\sigma}(i\omega_n )$ is an effective transverse susceptibility matrix, and $\chi^{\sigma,-\sigma}_0(i\omega_n )$ is the bare transverse susceptibility~\cite{ka.li.99}. 
The Matsubara frequencies are the complex energies $i\omega_n=i(2n+1)\pi T$, where $n=0,1,2,...$ and $T$ is the  temperature, and  $m$ corresponds to the magnetic interaction channel~\cite{bi.sc.89,ka.li.99}. The local Green's functions as well as the electronic self-energies are spin diagonal for collinear magnetic configurations. In this approximation the electronic self-energy is calculated in terms of the effective interactions in various channels. 
The particle-particle contribution to the self-energy  was combined with the Hartree-Fock and the second-order contributions \cite{ka.li.99}. 
To ensure a  physical transparent description the combined particle-particle self-energy is presented by Hartree,
$\Sigma^{(TH)}(i\omega_n )$, and Fock, $\Sigma^{(TF)}(i\omega_n )$, type contributions:
$\Sigma(i\omega_n ) = \Sigma^{(TH)}(i\omega_n )+ \Sigma^{(TF)}(i\omega_n ) + \Sigma^{(ph)}(i\omega_n )$,
where the particle-hole contribution $\Sigma^{(ph)}$ reads:
\begin{equation}\label{selfph}
\Sigma_{12 \sigma} ^{(ph)}(i\omega_n ) = \sum_{34 \sigma^{\prime}} W_{1342}^{\sigma
\sigma^{\prime}}(i\omega_n ) G_{34}^{\sigma^{\prime}}(i\omega_n ).
\end{equation}
A Pad\'e~\cite{vi.se.77,os.ch.12} analytical continuation is employed to map the self-energies from the Matsubara frequencies onto real energies, as required in the transmission calculation.
Since the static contribution from correlations is already included in the LSDA, so-called ``double-counted'' terms must be subtracted. In other words, those parts of the DFT expression for the total energy that correspond to the interaction included in the Hubbard Hamiltonian has to be subtracted.
To achieve this, we replace
$\Sigma_{\sigma}(E)$ with $\Sigma_{\sigma}(E)-\Sigma_{\sigma}(0)$
\cite{li.ka.01} in all equations of the DMFT procedure \cite{ko.sa.06}.
Physically, this is related to the fact that DMFT only adds {\it dynamical} correlations to the LSDA result. For this reason, it is believed that this kind of double-counting subtraction is more appropriate for a DMFT treatment of metals than the alternative static Hartree-Fock (HF) subtraction~\cite{pe.ma.03}.

The analytically continued self-energy (obtained in the DMFT), within the FPLMTO basis set is transfered into the multiple-zeta basis of {\sc siesta} according to the basis transformation presented in Ref.~\onlinecite{ch.mo.15}. 
With the corresponding self-energy we compute the interacting Green's function and use the latter in the expression for the transmission:
\begin{eqnarray}
\mathbf{G}^\sigma_{DMFT}(\mathbf{k}_\parallel,E) &=& \left[
\epsilon^+\mathbf{S}(\mathbf{k}_\parallel) - \mathbf{H}^\sigma(\mathbf{k}_\parallel)
-\mathbf{\Sigma}_{L}^{\sigma}(\mathbf{k}_\parallel,E)
-\mathbf{\Sigma}_{R}^{\sigma}(\mathbf{k}_\parallel,E)
-\mathbf{\Sigma}_{DMFT}^{\sigma}(E) 
\right]^{-1},
\label{Green_add_MB} \\
T^{\sigma}_{DMFT}(\mathbf{k}_\parallel,E) &=& {\rm Tr} \left[
\mathbf{\Gamma}_{L}^\sigma (\mathbf{k}_\parallel,E) \mathbf{G}_{DMFT}^{\sigma\dagger} (\mathbf{k}_\parallel,E)
\mathbf{\Gamma}_{R}^\sigma (\mathbf{k}_\parallel,E) \mathbf{G}_{DMFT}^{\sigma}(\mathbf{k}_\parallel,E)\right].
\label{landauer_dmft} 
\end{eqnarray}
The self-energy acts as a spin- and energy-dependent potential whose imaginary part produces a broadening of the single-particle states due to the finite electron-electron scattering lifetime.
Note that this is an approximation, since it neglects vertex corrections due to in-scattering
processes~\cite{oguri.01,me.wi.92}, which in general increases the conductivity.
Since we are not performing  fully self-consistent transmission calculations our $\mathbf{\Gamma}_{L/R}^\sigma (\mathbf{k}_\parallel,E)$ remain at the LDA/GGA level.
We are not aware of any method that allows to perform fully self-consistent calculations of $\mathbf{\Gamma}_{L/R}^\sigma (\mathbf{k}_\parallel,E)$ for realistic materials.

\subsection{Computational details}
\label{sec:compdetails}

Bulk NiMnSb crystallize in the face-centered cubic (fcc) structure with three atoms per unit cell, with positions Ni ($0$,$0$,$0$), Mn ($\frac{1}{4}$,$\frac{1}{4}$,$\frac{1}{4}$), and Sb ($\frac{3}{4}$,$\frac{3}{4}$,$\frac{3}{4}$). In the FPLMTO calculations, the radii of the non-overlapping muffin-tin spheres were set to 2.02 a.u. (Ni), 1.99 a.u. (Mn), 2.21 a.u. (Sb), and 2.38 (Au) respectively. For Ni and Mn the $3s$, $3p$, $4s$ and $3d$ electrons were treated as valence, while for Sb the $5s$, $5p$ and $4d$ electrons were considered to be valence electrons. For the Au leads $6s$ and $5d$ where treated as valence. For the bulk case a \textbf{k}-point mesh of size $32 \times 32 \times 32$ was employed with a Fermi-Dirac smearing function for the Brillouin zone integrations. 
The angular momentum cutoff for the charge density was chosen as $l_{max}=8$. Three kinetic energy tails corresponding to $0.3$, $-2.3$ and $-1.5$ Ry were employed. The exchange-correlation potential was approximated using the Perdew and Wang parametrization for the LSDA~\cite{pe.wa.92}. By direct computation, spin-orbit effects were found to be negligible for the quantities of interest, and hence we only kept the scalar-relativistic terms in our calculations.
The Matsubara sums were truncated after 1024 frequencies and the temperature  was set to $T=2\,$mRy.

\section{Results}
\label{sec:results}
This section presents the results for the electronic structure of bulk NiMnSb, and different terminations of NiMnSb(001) surfaces. We compare our results with previously reported studies at the level of DFT (LDA/GGA), and present novel results using DMFT. These results for the NiMnSb(001) surfaces are then compared with the NiMnSb(001)  interface to Au and for the latter ones the corresponding transmission coefficients are discussed. 
  
\subsection{Electronic structure of bulk NiMnSb, NiMnSb(001) surfaces and interfaces}
\label{sec:elstruc}

A large number of calculations using DFT methods are present  for bulk NiMnSb~\cite{gr.mu.83,ga.de.02,ka.ir.08} and its surface states~\cite{je.ki.01,le.ma.06}. 
According to these results, the bulk minority spin indirect gap is formed between the $\Gamma$ and $X$ points of the Brillouin zone corresponding to the fcc unit cell.
The minority spin bands consists of completely occupied Sb $p$-states~\cite{gr.mu.83,ka.ir.08}, while the bonding and anti-bonding $d$-hybrids of the Mn and Ni atoms are separated by the half-metallic gap. 
The bonding states have most of their weight at the Ni atom and the anti-bonding states at the Mn atom leading to very large localized spin moments at the Mn atoms~\cite{pl.sc.99}. The total spin moment follows the Slater-Pauling behavior as shown in Ref.~\onlinecite{ga.de.02}, being exactly $4\,\mu_B$. This is mainly determined by the ferromagnetic alignment of the large Mn spin moment ($3.72\,\mu_B$) and a small Ni one ($0.28\,\mu_B$).
Including spin-orbit coupling leads to a partially filled minority spin gap~\cite{ma.sa.04}. The majority spin DOS around the Fermi level changes only marginally, and the material remains essentially half-metallic with a polarization of the DOS of about $99\%$~\cite{ma.sa.04}. 
DMFT calculations, on the contrary, show the presence of NQP states, just above the Fermi level~\cite{ch.ka.03}.

Terminations of NiMnSb(001) surfaces are possible either with a Ni or a MnSb layer. 
These interfaces were studied previously~\cite{le.ga.05,je.ki.01} and great attention has been given to the relaxation effects. The results of the study of the relaxation effects upon the first two interface layers can be briefly summarized as follows: 
a) in the case of the Ni termination, almost no buckling or relaxation of the MnSb subsurface layer was observed, the distance between the top Ni layer and the subsurface layer was reduced by around $10\%$; 
b) in the case of the MnSb termination, the Mn atom at the surface layer moves inwards and the Sb atom outwards: the distance between the Mn surface atom and the Ni subsurface layer is contracted by $3.5 \%$ and the distance between the Sb surface atom and the Ni subsurface layer is expanded by $7.3 \%$. 
Relaxation effects upon the DOS have also been discussed, and proven that no significant relaxation effects are seen in the DOS~\cite{gala.02,ri.no.00,bo.ko.00,bo.ko.01}. 
Surface states were shown to be strongly localized at the surface layer, as in the subsurface layer there are practically no states contributing to the DOS inside the gap~\cite{le.ga.05,je.ki.01}. 
The Ni-terminated surface states are localized to particular surface layers and lead to flat dispersions.
Accordingly the DOS of these surface states are much more pronounced compared to those of MnSb terminated surface states and effectively destroy the minority spin gap. 

\subsubsection{The electronic structure of bulk NiMnSb}
\label{sec:bulk_001}
\begin{figure}[h!]
    \centering   
\includegraphics[width=\textwidth,clip=true]{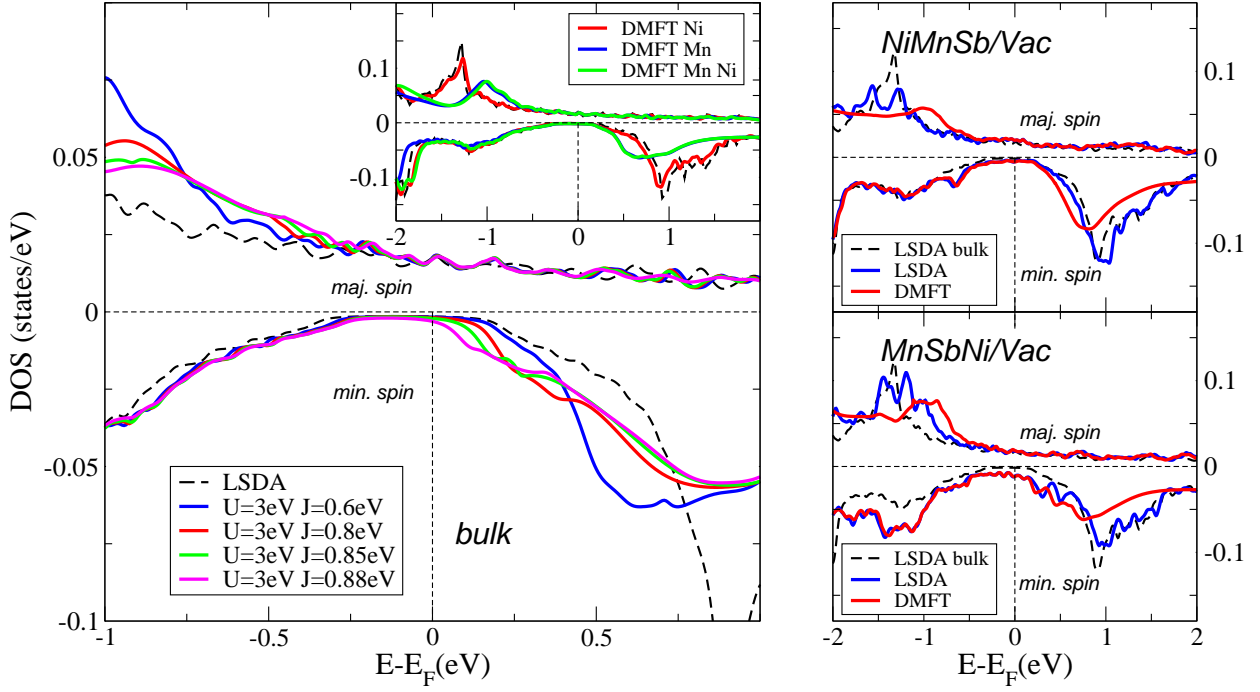}
\caption{(Left) Total DOS for bulk NiMnSb. The black dashed lines indicates the LSDA results. Solid lines represents LSDA+DMFT results including Hubbard corrections ($U=3\,$eV) and various Hunds exchange couplings (only Mn treated as correlated). The inset shows the DOS for $U=3\,$eV and $J=0.6\,$eV. The legend indicates which species are treated as correlated.  
(Right) Results for the NiMnSb (001) surface. On top MnSb termination, bottom Ni termination. Solid lines LSDA (blue) and DMFT (red).
The following parameters were used in the DMFT calculations: $U=3\,$eV, and $J=0.6\,$eV.}
\label{Fig:1}
\end{figure}

In figure Fig.~\ref{Fig:1} we present the LSDA and LSDA+DMFT results for the total DOS calculations of the bulk NiMnSb (left) and the slab NiMnSb/Vac (right).
The DOS is normalized in a way that the integrated DOS gives one at the Fermi-level. 
Fig.~\ref{Fig:1}, left panel, represents the DOS in the energy window $E_F \pm 1\,$eV around the half-metallic gap. In its inset the total DOS is presented in a larger energy range $E_F \pm 2\,$eV. To facilitate the comparison, the same energy range is used for the total DOS of the slab calculations (Fig.~\ref{Fig:1}, right panel).  

The LSDA results for the bulk NiMnSb correspond to the case in which no $U$ and $J$ corrections are applied. 
A minority spin gap of size  $\approx 0.5\,$eV is visible in the DOS. 
We name the energy range where the minority spin DOS develops a gap in the LSDA calculation the gap region. 
This result is in good agreement with the experiment of Kirillova et al.~\cite{ki.1995}, who analyzed infrared spectra and estimated a gap width of $\approx 0.4\,$eV. 
Previously published band structure calculations report  similar band gaps in comparison to our result~\cite{gr.mu.83,gr.mu.83,ga.de.02,ya.ch.06,ka.ir.08}.
The minority spin gap is formed within the $d$-band manifold between the bonding ($t_{2g}$) and anti-bonding ($e_{g}$) states. Above the Fermi level, Mn $d$-orbitals dominate the gap edge.   

In order to study the electron-electron correlation effects and its influence upon the minority spin gap we performed calculations with different values of the local Coulomb interaction and exchange parameters  up to $U=3\,$eV and $J=0.88\,$eV. 
When Mn sites are treated as correlated (main graph) we observe significant changes in the DOS: peaks are shifted and simultaneously the gap is filled just above the Fermi level.
These results are in agreement with previous LSDA+DMFT calculations~\cite{ch.ka.03}.
The origin of the in-gap states is connected with "spin-polaron" processes: the spin-down low-energy electron excitations, which are forbidden for half-metallic ferromagnets in the one-particle picture, turn out to be possible as superpositions of spin-up electron excitations and virtual magnons~\cite{ka.ir.08,ka.ed.92,ir.ka.05,ir.ka.06,ch.ka.03}.
The density of these states vanishes at the Fermi level $E_F$ at zero temperatures, and increases at the energy scale of the order of a characteristic magnon frequency~\cite{ka.ir.08,ka.ed.92,ir.ka.05,ir.ka.06,ch.ka.03}.
Recently, the density of NQP states has been calculated from first principles for semi-Heuslers~\cite{ch.ka.03,ch.ar.06}, Heuslers~\cite{ch.ar.09,ch.sa.08} and zinc-blend half-metals~\cite{ch.ka.05,ch.ma.06} and heterostructures~\cite{be.ho.11,ch.le.11}.

For bulk NiMnSb we have studied the formation of the NQP state as a function of the interaction parameters. 
In particular we observe that the Hund’s coupling $J$, the intra-atomic exchange energy, crucially influences the DOS. Increasing the value of $J$ in the range of $0.6\,$eV up to about $0.88\,$eV, the NQP state (broad shoulder) is  shifted towards the Fermi level. Such behavior can be captured within a $s$-$d$ type model for the electron-magnon interaction~\cite{ir.ka.05,ir.ka.06,ka.ir.08}, and was shown to be valid also for the Hubbard model, within the DMFT approximation~\cite{ch.ka.03}. In the former case, the corresponding change in the spectral density (DOS) is caused by 
a term, proportional to the real part of the self-energy, shifting the quasiparticle energies.
There is a second term that arises from the branch cut of the self-energy and which describes the incoherent, ``non-quasiparticle'', contribution (proportional to the imaginary part of the self-energy). All these effects are visible in  Fig.~\ref{Fig:1}, consequently the formation of NQP states in half-metals can be explained as the low energy physics of electron-magnon interaction. 

In the inset we show a comparison of the LSDA+DMFT results when both, or only one of the $d$-electron  subsystems is treated as correlated. Adding the Hubbard corrections only to Ni, we find only minor changes with respect to the LSDA results. Considering both Mn and Ni as correlated, the results do not differ much from the case in which only Mn is treated as correlated.
The predominant correlation effect of Ni is expected in a region of higher binding energies~\cite{li.ka.97}.
Since correlation effects originating from the Ni site affect the DOS only slightly in the energy range close to $E_F$ we treat Ni as uncorrelated in the following surface and interface calculations.

\subsubsection{NiMnSb (001) surface terminations.}

On the right hand side of Fig.~\ref{Fig:1} the total DOS for (NiMnSb)$_2$ with two possible (001) surface terminations are shown. 
We investigated the MnSb-termination and the Ni-termination on both surfaces of the slab. 
The overall shape of the LSDA-DOS is similar to the bulk one for both surface terminations, except the DOS in the gap region: we find no longer a gap in the minority spin DOS, instead in-gap localized surface states are formed. 
The in-gap states for Ni-terminated surfaces are more intense then the ones for MnSb-terminated surfaces. 
Our LSDA results are in agreement with previous calculations~\cite{le.ga.05,je.ki.01}.
Independent of the surface termination, the peak in the bulk-DOS (around $E-E_F\approx -1.5\,$eV) of hybrid Ni and Mn bonding states of e$_g$ character are reduced in magnitude and split. The splitting of this peak is determined by the symmetry reduction from the cubic to tetragonal symmetry.
As a consequence of splitting the weight is redistributed and the overall magnitude of DOS is decreasing.
For the Ni-terminated structure the minority spin DOS has more weight in the gap region in comparison to the MnSb-terminated structure, as visible in the lower right graph of Fig.~\ref{Fig:1}. 
 
The LSDA+DMFT results for the DOS follow the general trend of the corresponding results for the bulk. The in-gap surface states are already present at the LSDA level. 
They are situated in the same energy range as the NQP states and make the NQP peak not discernible in the DOS. The only visible correlation effect remains the shift of the one-particle states. 
In comparison with the LSDA spectra the minority-spin unoccupied and the majority-spin occupied DOS shifts towards the Fermi level. 

\subsubsection{Electronic structure at the NiMnSb/Au interface}

\begin{figure}[h]
\includegraphics[width=0.45\columnwidth]{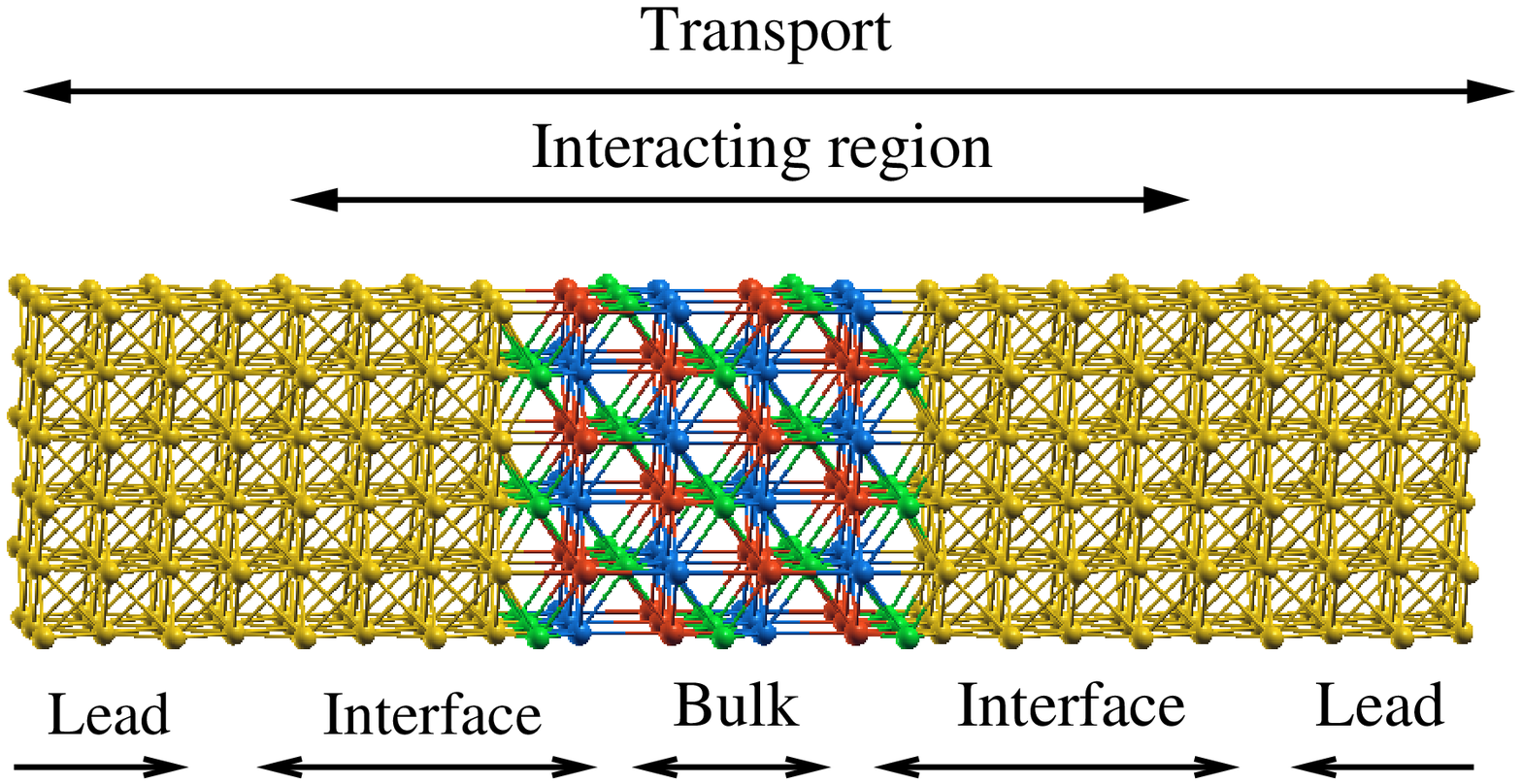} 
\includegraphics[width=0.45\columnwidth]{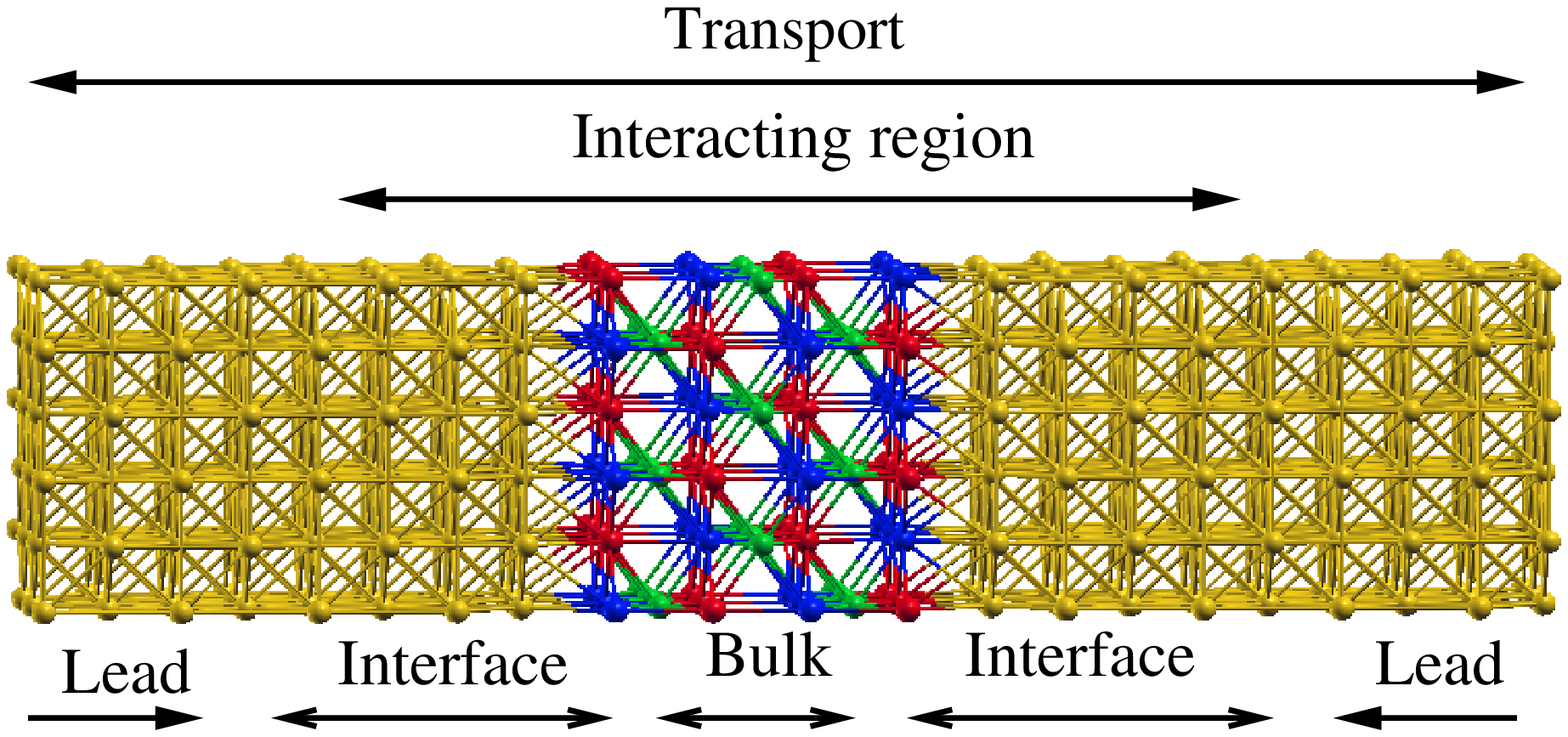}
\caption{Schematic representation of the supercell used in the calculations. (Left) Ni-terminated (001) interface. (Right) MnSb-terminated interface. We indicate the regions where the electronic structure is assumed similar to that of the bulk as well as the size of the cells used in transport and DFT calculations, respectively. Atomic color code: Ni (green), Mn (blue), Sb (red), Au (yellow)}
\label{fig:struc}
\end{figure}

We consider the contact geometry as shown in Fig.~\ref{fig:struc}. The geometry is similar to surface structures in which the interface to vacuum is replaced by metallic gold leads. 
For the NiMnSb/Au interface the contact is taking place between the gold layer and one of the two possible layer terminations: The Ni-terminated structure is shown on the left part of Fig.~\ref{fig:struc} and the MnSb-terminated structure is shown on the right part of Fig.~\ref{fig:struc}.

\begin{figure}[h!]
\includegraphics[width=\textwidth]{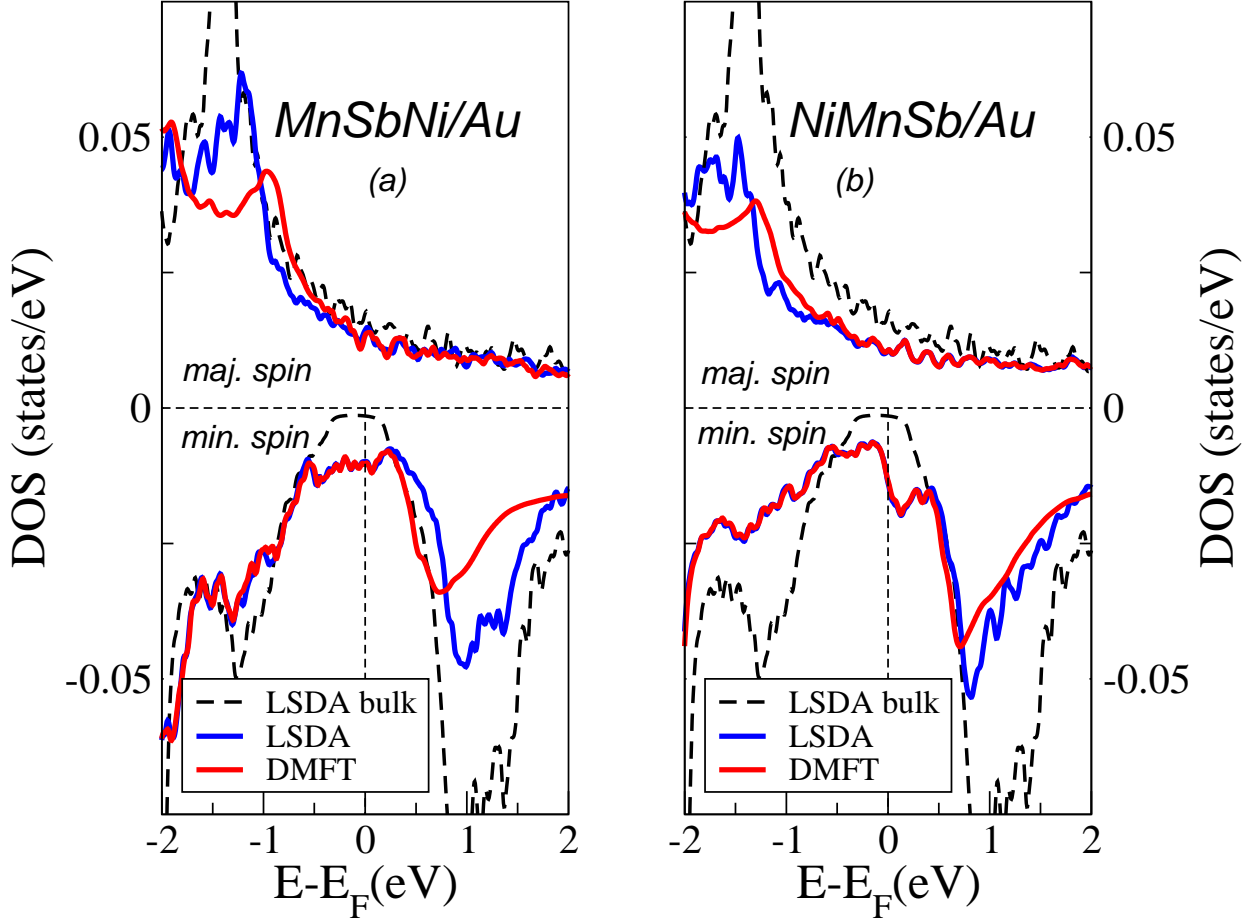}
\caption{Total DOS for the Au--NiMnSb--Au slab. (a) MnSb termination. (b) Ni termination.  The solid-blue line indicates the LSDA DOS and the solid-red line indicates the LSDA+DMFT DOS. The LSDA+DMFT results were obtained for $U=3\,$eV and $J=0.6\,$eV.}
\label{Fig:2}
\end{figure}

In the upper and lower parts of Fig.~\ref{Fig:2} we present the majority and  respectively the minority total DOS for the case in which two NiMnSb units in the interacting region are considered. With blue and red lines we denote the LSDA and respectively the DMFT results for the DOS. In the left and right panels the results for the Ni- and MnSb-terminated interfaces are presented respectively. For comparison we have plotted in Fig.~\ref{Fig:2} with dashed lines the bulk NiMnSb DOS with a gap of the size $\approx 0.5\,$eV in the minority spin channel. 

For both interfaces one can clearly see that states appear inside the minority spin gap region in LSDA and LSDA+DMFT calculations. These states are most likely localized near the interface region, similarly to the case of the NiMnSb(001) surface~\cite{le.ga.05}.
However, because of the presence of Au layers, a stronger weight of the in-gap-states is obtained. 
Note that no significant differences between correlated and non-correlated calculations are seen in the close vicinity of $E_F$ for both spin channels.
For the Ni-termination the minority-spin gap is populated with an almost constant DOS (Fig.~\ref{Fig:2}, left panel).
For the MnSb-termination minority-spin DOS just above the Fermi level reaches a maximum at about $0.1\,$eV (Fig.~\ref{Fig:2}, right panel). No such maximum value of the DOS is seen in the vicinity of $E_F$ for the surface calculation (see Fig.~\ref{Fig:1}). 
The origin of this maximum in the minority-spin DOS, just above the Fermi level, can be attributed to the hybridization between the interface Au $s$- and the Sb $s$-orbitals of the interface MnSb layer. For Ni-termination no such hybridization can take place, therefore the constant DOS in the minority spin channel (Fig.~\ref{Fig:2}, left panel) represents in fact the interface Ni states. 

Correlation effects are visible only outside the gap region for both spin channels. For both terminations, changes in the DOS take place below $E_F$ for the majority-spin electrons and below $E_F$ for the minority-spin electrons. These changes represent correlation-induced spectral weight shifts and are consequences of the negative slope of the real part of the self-energy. 

The signature of many-body features in the spectral function of half-metallic ferromagnets is the existence of NQP states. NQP states are generic many-body features for all ferromagnetic materials, however in half-metals their weight and position is visible because of the existence of the half-metallic gap. 
In bulk NiMnSb the NQP states are visible in Fig.~\ref{Fig:1}, their weight depends significantly on the strength of the exchange coupling $J$, and they inherit the $d$-orbital character of the Mn atoms~\cite{ka.ir.08,ch.ka.03}.
The non-quasiparticle contributions to the DOS, originates from the imaginary part of the minority spin self-energy~\cite{ka.ir.08,ir.ka.05,ir.ka.06}:
which produces a branch-cut in the corresponding minority spin channel Green's function. 
A similar result exists for the surface states~\cite{ka.ed.92} and generalized for arbitrary inhomogeneous cases~\cite{ir.ka.06}, and is therefore also valid for interfaces. 
In order that the NQP states become visible in the spectral function (DOS), \textit{interface} half-metallicity has to happen, with the half-metallic gap formed between correlated orbitals. In the present case of the NiMnSb/Au interface, the half-metallic character is lost because of the interface hybridization in the case of MnSb-termination or because of extended surface Ni states in the Ni-terminated setup. 
Consequently NQP states cannot be distinguished clearly in the DOS, since they are overlapped by the interface states.

In situations, where the gap region in the minority-spin channel is filled by surface states, the minority spin tunneling process is mediated by the surface states only~\cite{ir.ka.06}. Although the NQP states are overlapped, in the DOS, their presence can be revealed through the tunneling current due to hybridization of bulk-states with surface or interface states~\cite{ka.ed.92,ir.ka.06}.

\subsection{Transport properties: the Au--NiMnSb--Au heterostructure setup}
\label{sec:trans}

Within the standard approach for ballistic transport calculations the system is partitioned into two semi-infinite parts (the leads) and the scattering zone. From a qualitative point of view the transmission probability is essentially determined by the scattering matrix connecting the states from the left lead to those in the right lead.
The structural setup consists of two gold leads in the Au(100) orientation that sandwich atomic planes in the pattern AB-AB-AB (A = Ni, B = MnSb) along the transport direction. 
The Au leads structure was build using the experimental value for bulk lattice parameter (4.08~\AA). 
The NiMnSb units are built on top of this structure and orient accordingly with the two possible terminations (see Fig.~\ref{fig:struc}). 
The region where the electron scattering occurs (interacting region) contains a few layers of the two substrates (the leads) and sequences of Ni and MnSb layers as indicated in Fig.~\ref{fig:struc}. The electronic structure of the NiMnSb units within the interacting region is different from that of the bulk, due to the presence of the  interface (see Sec.~\ref{sec:elstruc}).
For thick layers (e.g., including three or more units of NiMnSb) we expect to find in the center of the interaction zone the electronic structure of the bulk NiMnSb. Therefore, for such thick layers, the scattering properties across the whole structure should reflect the electronic structure of the bulk. In these cases, the half-metallic character of the electronic structure should be reflected in the transmission probability.

\subsubsection{Transmission computed within GGA}
In the following we analyze the results of the transmission in view of the changes brought by the presence of the interfaces and in view of the electronic structure results.
The basis set used in the {\sc siesta} and {\sc smeagol} calculations is of ``double-zeta with polarization'' (DZP) quality. The ``energy shift'' parameter which allows to control the extent of basis functions on different atoms was taken to 300~meV, that resulted in basis functions with a maximum extent of $6.1\,a_0$ (Au), $6.0\,a_0$ (Ni), $6.4\,a_0$ (Mn) and $4.9\,a_0$ (Sb) where $a_0$ is the Bohr radius. The PBE-GGA~\cite{pe.bu.96} functional of DFT has been used and no relaxation has been considered.
The spin-resolved transmission probability, $T_\sigma(E)$, is obtained from the $\mathbf{k}$-dependent transmission, by integrating over all $\mathbf{k}_\parallel$-points, so that $T_\sigma(E)=\frac{1}{\Omega_\mathrm{BZ}}\int_\mathrm{BZ}d\mathbf{k}_\parallel T_\sigma(\mathbf{k}_\parallel,E)$. The $(\mathbf{k}_\parallel,E)$-dependent transmission Eq.~(\ref{landauer}) or~(\ref{landauer_dmft}) is obtained from the matrix product of the hybridizations $\mathbf{\Gamma}_{L/R}^\sigma (\mathbf{k}_\parallel,E)$ and the retarded/advanced Green's functions $\mathbf{G}^{\sigma}(\mathbf{k}_\parallel,E)/\mathbf{G}^{\sigma\dagger}(\mathbf{k}_\parallel,E)$.
While the Green's functions encodes the electronic structure of the interacting region, the hybridization function carries the information about the hopping of electrons into and out of the interacting region.

\begin{figure}[h]
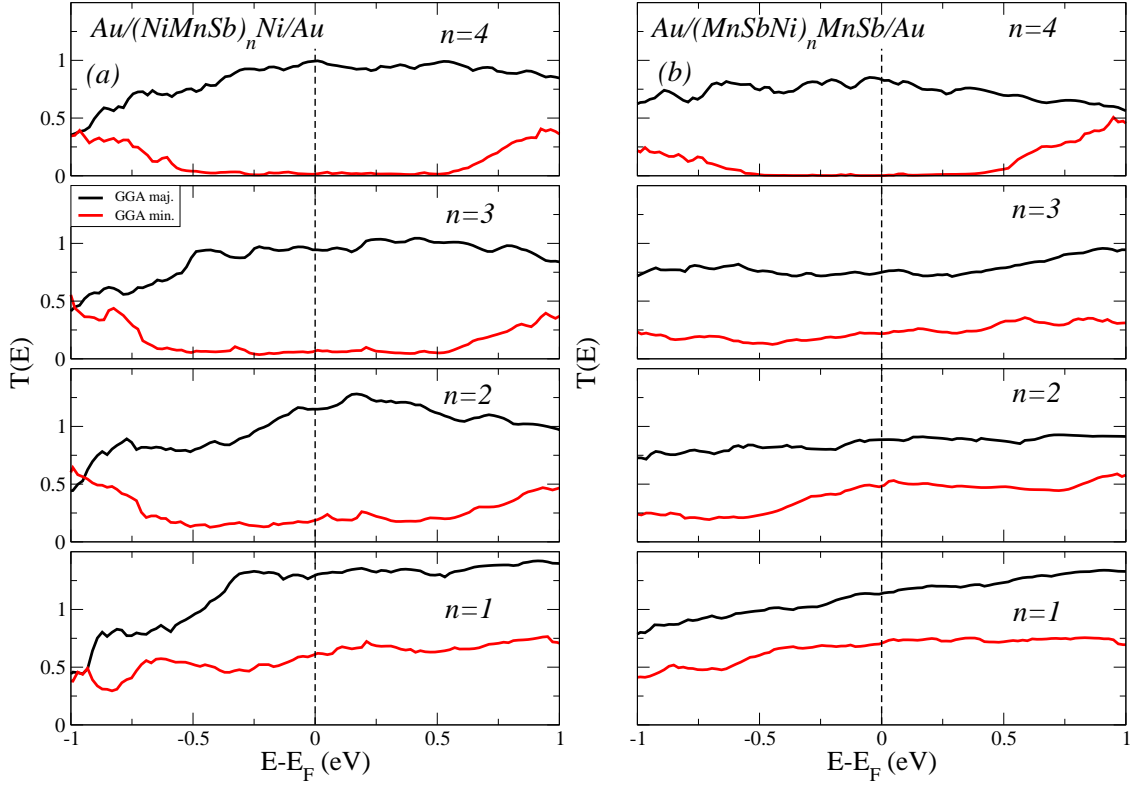

\includegraphics[width=0.45\columnwidth,clip=true]{Fig4a.eps}
\includegraphics[width=0.45\columnwidth,clip=true]{Fig4b.eps}
\caption{Evolution of DFT (GGA) spin-resolved transmission as a function of the number  $n$ of (NiMnSb)$_n$ units: (a) Ni terminated interfaces and (b) MnSb terminated interfaces. The black solid lines represent the majority spin transmission and the red solid lines indicate the minority spin transmission.
}
\label{Fig:fig4}
\end{figure}

In Fig.~\ref{Fig:fig4} we display the total spin-resolved transmission probability computed with the GGA. It can be immediately seen that the transmission displays considerable spin polarization. For a number of units larger than three ($n \geq 3$) the spin polarization at $E_F$ is above 90$\%$ for both terminations.
Interestingly, for both the Ni- and the MnSb-terminated structure, the ``bulk-like behavior'' (i.e., ``half metallic'' transmission) is present already for a number of $n=4$ NiMnSb units. 
For a smaller number of NiMnSb units a possible direct lead-to-lead conduction channel may form 
which determines the non-zero transmission for the minority spin electrons. 
The transmission in the majority spin channel (solid black lines) is slightly smaller for the  MnSb-terminated structure in comparison with the Ni-terminated structure. 

\subsubsection{Transmission computed within GGA+DMFT}

\begin{figure}[h!]
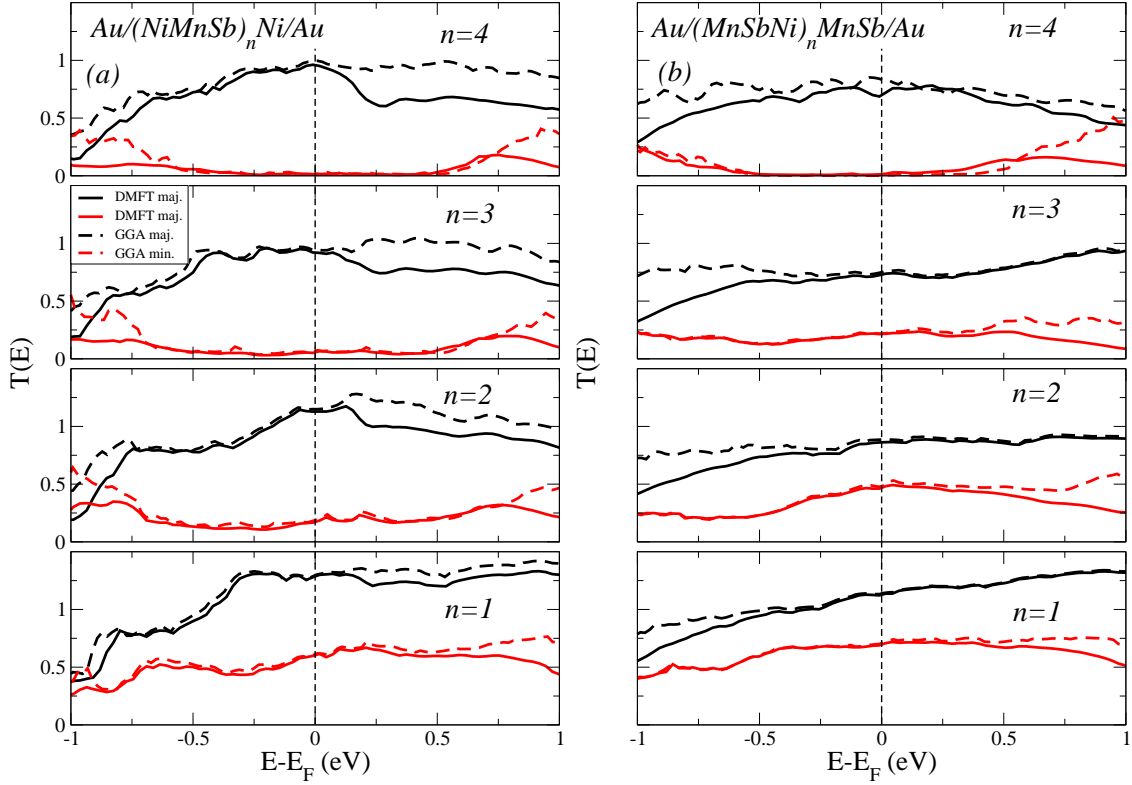

\includegraphics[width=0.45\columnwidth,clip=true]{Fig5a.eps}
\includegraphics[width=0.45\columnwidth,clip=true]{Fig5b.eps}
\caption{Spin-resolved transmissions. The dashed lines represent the GGA results and the solid lines represent the GGA+DMFT results. 
The black colored line denotes the majority spin transmission and the red colored line denotes the minority spin transmission:
(a) the transmission through Ni-terminated structure (b) the transmission through MnSb-terminated structure.
}
\label{Fig:fig5}
\end{figure}

In the following we analyze the results of transmission in the view of the changes brought by the presence of  electronic correlations.
The GGA+DMFT transmissions for Ni- and MnSb-terminated structures are shown in Fig.~\ref{Fig:fig5}.
On the overall energy scale the GGA+DMFT and GGA transmissions have a similar energy dependence, however the magnitude of the transmission is reduced because of electronic correlations.

For the Ni-terminated structure (left side of Fig.~\ref{Fig:fig5}) changes induced by the presence of electronic correlations are visible above the Fermi level in both spin channels. 
For the MnSb terminated structure, changes induced by electronic correlations are spin selective. For the majority spin (black line, Fig.~\ref{Fig:fig5} right column) a more significant reduction in transmission is seen below $E_F$.
This is a consequence of the larger/smaller weight of the imaginary part of the Mn self-energy below/above $E_F$. 
A completely opposite effect is seen for the minority spin electrons (red line). In this spin channel the imaginary part of the manganese self-energy has a smaller/larger weight below/above $E_F$. 

The depletion in the transmission is determined by the decrease in the coherence of the scattered wave-function across the heterostructure. 
The inverse of the non-zero imaginary part of the self-energy corresponds to the  finite lifetime of the quasiparticle and broadens the spectral function (DOS).
The above statements are valid for all layers in the scattering region that are subject to electronic correlations.
We found that the correlation-induced changes of the transmission is almost independent of $n$.
For $n\le 3$ the tunneling is influenced considerably by lead-to-lead direct tunneling and signatures of the electronic structure of NiMnSb are not clearly visible. For $n=4$ a significantly large  spin-polarization of the transmission sets in, signaling the importance of half-metallicity in the scattering region. Away from the Fermi level a hump is formed at around $0.4\dots 0.6\,$eV in the minority spin-channel. 

\begin{figure}[h]
\includegraphics[width=\columnwidth]{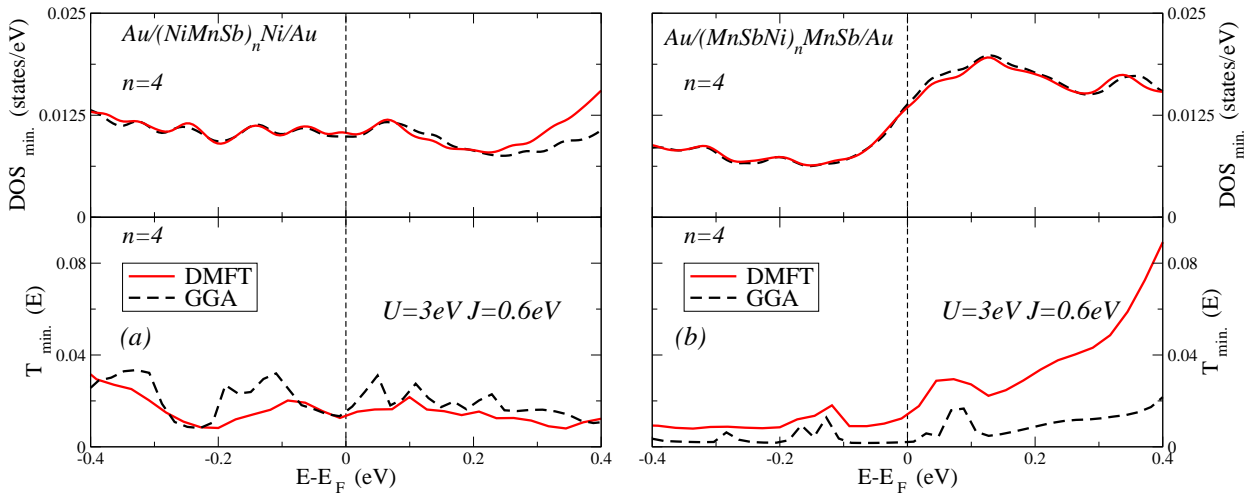}
\caption{Comparison of the minority spin DOS$_{min}$ (upper panel) and the minority spin transmission T$_{min}(E)$ (lower panel). (a) Ni-terminated structure (b) MnSb terminated structure.}
\label{Fig:fig6}
\end{figure}

Surface and interface electronic structure calculations in
Sec.~\ref{sec:bulk_001} show that the minority spin gap is filled by surface states, that are more important than many-body effects (the NQP features in Fig.~\ref{Fig:1}). A more quantitative analysis may result from the direct comparison of the DOS and the transmission around $E_F$. 
In the upper part of Fig.~\ref{Fig:fig6} we show the results of the DOS for the heterostructures with four units NiMnSb in the scattering region, Ni-terminated (left) and MnSb-terminated (right). 
In the lower part of Fig.~\ref{Fig:fig6} the transmissions are presented for the corresponding structures. We focus on a narrow energy range around the Fermi level, $E_F \pm 0.4\,$eV. For the Ni-terminated structure, no change in the DOS is seen in the range $-0.4$ to $0.2\,$eV. On the other hand the transmission (lower part) is reduces in the entire energy range. For the MnSb termination, no significant change in DOS in seen in the $E_F \pm 0.4\,$eV energy range. On the other hand the transmission is enhanced by electronic correlation. This might be understood as tunneling assisted by quasiparticle states or by non-quasiparticle states in the scattering region~\cite{ir.ka.06}. 

The modification of the transmission amplitude caused by the finite real part of the self-energy can be understood as a renormalization of the one-particle states within the scattering region. 
A second mechanism that modifies the transmission amplitude is given by the incoherent part of the minority spin self-energy caused by hybridization of surface states with NQP states. 
In order to resolve the true nature of the changes in the minority spin transmission 
one could investigate the temperature and bias dependence of the tunneling current. It has been shown previously that the quasiparticle contribution to the tunneling current can even be more pronounced in comparison to the NQP peak in the DOS~\cite{ir.ka.06}. It is worth to emphasize that in the Landauer-B\"uttiker approach the many-body equilibrium self-energy is used to compute the transmission in a first step which is then used to compute the tunneling current in a second step. This implies that bias and temperature dependence of the tunneling current is determined solely by the Fermi-Dirac distribution containing these parameters. 
Hence the Landauer-B\"uttiker formalism explicitly excludes genuine correlation effects on the bias dependence of the tunneling current by construction. We think that the hump around $0.4\dots 0.6\,$eV (seen in Fig.~\ref{Fig:fig5}) in the minority-spin  transmission can be considered to be a precursor of the bias dependence of the tunneling current.
In order to address the implications of the hump in the minority spin transmission on the tunneling current in a more rigorous way one would need to extend our present transport formalism to a full charge self-consistent calculation employing the NEGF including many-body corrections on the level of DMFT.  

\section{Conclusion}
\label{sec:conc}

The origin of the bandgap in half-Heusler alloys (XYZ) is the hybridization between the $3d$ states of the X and Y elements (here: X = Ni,Y = Mn). 
The minority spin gap is formed between the bonding ($t_{2g}$) and anti-bonding ($e_{g}$) states. 
It is believed that the half-Heusler NiMnSb alloy with its bandgap of $\approx 0.5\,$eV is promising for the development of high-performance magnetoresistive devices, because of the suppression of thermal activation of the electrons. 
Previous studies have demonstrated that the half-metallic properties of half-Heusler alloys can be very easily degraded by various factors.
In bulk NiMnSb the many-body states induced by the interaction are formed just above the Fermi level~\cite{ch.ka.03,ch.ar.06,ch.ar.09,ka.ir.08}. 
It has been shown that the current-perpendicular-to-plane giant magneto-resistance ratio is considerably reduced due to sample defects~\cite{caba.98}. 
This indicates that a precise control over the sample purity and structure is highly desirable.

Electronic structure calculations for bulk and (001) surfaces with different terminations are  discussed extensively in the literature~\cite{gr.mu.83,ga.de.02,ka.ir.08,je.ki.01,le.ma.06}.
While DMFT results for bulk NiMnSb have been obtained previously~\cite{ch.ka.03,ch.ar.06,ch.ar.09,ka.ir.08}, no results including DMFT for the (001) surface and Au-capped NiMnSb layers have been reported before. 
In this work we study the effects of local electronic interactions upon the transmission across the NiMnSb layers sandwiched between gold leads. 
Electronic structure and transport calculations are presented using LSDA/GGA and DMFT extension.
The presence of the Au-leads bring $s$-type orbitals in the vicinity of the Ni- or MnSb-terminating layers. 
It is expected that the differences between the two interface terminations stem from the hybridization of Au $s$-states at the interface  with Sb $s$-states. 
In the case of MnSb-termination a strong hybridization is found.
Similar effects are less visible for the Ni-terminated geometry. 
In the latter case Sb $s$-states in the MnSb sublayer hybridize weaker with Au $s$-states. 

Concerning the results for transmissions, the general tendency is that electronic correlations reduce the magnitude of the transmission in the vicinity of the Fermi level. 
For the Ni-terminated structures the visible change happens above the Fermi level in both spin channels. 
For the MnSb-termination, a clear spin selectivity is obtained: majority spin transmission is diminished below $E_F$, while minority spin transmission is enhanced above $E_F$. 
It is interesting to note that the interface states cover the many-body induced non-quasiparticle states. 
Although the spin polarization of the density of states is considerably reduced (see Fig.~\ref{Fig:2}), the transmission polarization is not significantly affected by correlations (see Fig.~\ref{Fig:fig5}). 
A very large degree of polarization of the transmission is obtained in the case of four NiMnSb units, while a reduction of spin polarization to a value of up to $\approx 90\%$ is obtained for $n \le 3$. 
We believe that this reduction is due to the direct transmission (over the scattering region) from and into the leads. One of the major findings in this article is a very high spin-polarization in the transmission despite the presence of electronic correlation effects.
 
\section*{Acknowledgement}
We are grateful to Ivan Rungger for stimulating discussions. 
The calculations were performed in the data center of NIRDIMT. Financial support offered by the Augsburg Center for Innovative Technologies, and by the Deutsche Forschungsgemeinschaft (through TRR 80) is gratefully acknowledged. 
CM thanks UEFISCDI for financial support through project PN-III-P4-ID-PCE-2016-0217.
The research reported in this publication was supported by funding
from King Abdullah University of Science and Technology (KAUST).

\bibliography{main.bib}
\end{document}